\documentclass[sigconf, nonacm]{acmart}
\usepackage{algpseudocode}
\newcommand\vldbdoi{XX.XX/XXX.XX}

\newcommand\vldbavailabilityurl{}
\newcommand\vldbpagestyle{plain} 

\begin{document}
\title{A Generative Caching System for Large Language Models}

\author{Arun Iyengar}
\affiliation{%
  \institution{Cisco Research}
}
\email{aki@akiyengar.com}

\author{Ashish Kundu}
\affiliation{%
  \institution{Cisco Research}
}
\email{ashkundu@cisco.com}

\author{Ramana Kompella}
\affiliation{%
  \institution{Cisco Research}
}
\email{rkompell@cisco.com}

\author{Sai Nandan Mamidi}
\affiliation{%
  \institution{Cisco Research}
}
\email{saimamid@cisco.com}

\begin{abstract}
Caching has the potential to be of significant benefit for accessing large language models (LLMs) due to their high latencies which typically range from a small number of seconds to well over a minute. Furthermore, many LLMs charge money for queries; caching thus has a clear monetary benefit. This paper presents a new caching system for improving user experiences with LLMs. In addition to reducing both latencies and monetary costs for accessing LLMs, our system also provides important features that go beyond the performance benefits typically associated with caches. A key feature we provide is generative caching, wherein multiple cached responses can be synthesized to provide answers to queries which have never been seen before. Our generative caches function as repositories of valuable information which can be mined and analyzed. We also improve upon past semantic caching techniques by tailoring the caching algorithms to optimally balance cost and latency reduction with the quality of responses provided. Performance tests indicate that our caches are considerably faster than GPTcache.
\end{abstract}

\maketitle

\pagestyle{\vldbpagestyle}


\section{Introduction}

Caching is critically important for improving performance for a wide variety of applications in distributed computing systems. 
Recently, large language models~\cite{zhao2023survey, chang2024survey} like ChatGPT~\cite{van2023chatgpt} have become commonly used. Both the monetary cost and latency for using large language models models can be quite high. The ability to cache content for large language models offers significant advantages. For one thing, latency for satisfying queries can be significantly reduced. Queries for large language models can take several seconds or even minutes to satisfy. The savings in latency which can be achieved by caching LLM content are significant and often much more than the latency which can be saved by caching conventional web content.

Another key aspect is that LLMs often charge money for providing content. This means that successfully caching LLM content can reduce the cost for accessing LLMs in addition to reducing latency. This is a significant advantage provided by LLM caching which is not provided by most other forms of caching. Since LLM caching can reduce the monetary cost of accessing LLMs, the benefits for successfully deploying LLM caching are compelling.

In this paper, we present GenerativeCache, which is a caching and enhanced client system for large language models. GenerativeCache goes beyond previous caches by using a technique we call {\it generative caching} in which caches can synthesize new responses from responses that the cache is already storing. With generative caching, a cache can respond to queries that are semantically different from ones that it has received in the past.

The content provided by LLMs can be valuable either in its own right or in combination with other material. A cache containing LLM content can be used as part of a knowledge base which can be queried and used in conjunction with other materials to provide answers to more complex queries than the original ones resulting in the cached content.

Our architecture integrates caching with an enhanced client which coordinates access to multiple LLMs. Users can easily test out multiple LLMs and issue requests using different parameter settings through the client. Caching is integrated with the client, allowing users to easily control caches and optimize performance. This results in better performance and user experiences.

Our key contributions include the following:
\begin{itemize}
    \item We have developed a technique known as generative caching in which a cache can synthesize answers from multiple cached responses. Generative caching allows caches to synthesize responses to queries not seen before. Multiple caches can cooperate to synthesize responses to queries.

    \item Determining semantic similarity of queries is a key part of caching for LLMs. We improve on past work in semantic caching by adaptively varying the semantic similarity parameters to improve performance and user experience for a variety of situations, such as:
    \begin{itemize}
        \item Optimally handling different types of content (e.g. text versus computer code).
        \item Reducing monetary costs of queries.
        \item Reducing latency for satisfying queries.
        \item Accommodating user preferences for cost, response time, and quality of responses
    \end{itemize}
    \item We analyze the performance of GenerativeCache. Our system is efficient, with GenerativeCache being about 9 times faster than GPTcache, the most widely known LLM cache. Performance overhead is dominated by the time to compute embeddings for cache queries.
    \item We provide an enhanced client for LLMs integrated with our caches. This enhanced client allows users to interact with multiple LLMs while minimizing monetary costs and response times for queries. It also allows users to easily configure cache policies and parameters.
\end{itemize}

Caching LLM responses is different from caching content in other domains such as traditional web caching of images and static HTML pages. One key difference is the need to consider semantic similarity of queries. From this aspect, caching of LLM content is similar to query-response caching which has been previously performed in some query-answering systems. However, there are other differences as well. For example, LLMs can provide different types of content, such as natural language text, computer code and images. Having different caching policies for different types of content is appropriate. Another issue is the monetary costs required for making requests to an LLM for users who are paying customers. Caching can directly reduce those costs, which presents a compelling reason for attempting to cache LLM responses. The cost savings are thus obvious. By contrast, most forms of web and query caching are primarily deployed to reduce latency and computational resources. Caching by itself will usually not substantially reduce the costs paid to the provider of the content.

Another issue is that the latency for accessing a large language model can be orders of magnitude higher than accessing a static web file for caching. Requests to LLMs can take several seconds to well over a minute for completion. This means that the latency reduction which can be achieved by successful LLM caching can potentially be considerably higher than the latency reduction achieved by conventional web caching. It also means that we have leeway to perform more sophisticated operations by a large language model cache to generate the best responses. Even if our LLM cache takes a bit longer to provide a response than a conventional web cache, the ratio of response time of the LLM to response time of a cache is still quite high. 
\section{Semantic Caching Techniques Used by GenerativeCache}

While the benefits that can be achieved by LLM caching are significant, there are also complexities in building LLM caches which go beyond those of conventional caches. One key point is the importance of semantic similarity. Conventional caches typically store data referenced by a key. A key-value store can be used wherein cached values are stored and identified by keys.

This approach is not sufficient for storing the results of natural language queries. If a key-value store is used to store responses to natural language queries with the queries themselves being used as keys, then the only time a cache hit will occur is when two syntactically identical queries are made. This will result in low hit rates because several different queries can have very similar semantic meanings. For example, consider the query, “What is an application-level denial of service attack?”. This query is semantically quite similar to the query, “I would like to learn about application-level denial of service attacks. Please explain what they are.”. These queries are syntactically quite different, despite their semantic similarity. The conventional caching approach, wherein answers to these queries are indexed by the queries themselves, would not recognize the similarity of these queries.

To get around this problem, semantic caching is used wherein queries are compared using semantic similarity. A common approach is to vectorize the queries using a model and compare the vectors to determine semantic similarity of the queries. A wide variety of models are available for performing embeddings~\cite{patil2023survey}, and new ones are constantly being produced.

Both monetary cost and runtime overhead are critically important for using models in a caching system. Some models charge a fee based on the number of requests. Since the request rates can be high, free models are preferable. Furthermore, latency and CPU overhead for calculating embeddings are key considerations. Latency can be high if communication with a remote party is required for performing embeddings. For example, OpenAI offers multiple models for calculating embeddings, including its text-embedding-3-small, text-embedding-3-large, and text-embedding-ada-002 models. These three models offer different trade-offs regarding cost and how well the models perform. Using the models incurs higher latency than a local model because OpenAI needs to be contacted to perform the embeddings. While it is possible to make several requests for embeddings to OpenAI in parallel, there is a monetary cost for the requests. Furthermore, there are limits to how many requests users may make to OpenAI based on their plans. Because of these reasons, it would be preferable to use a high-performing model which can compute embeddings locally and without paying a third party for requests to embed data within vectors.

GenerativeCache supports multiple embedding models, as there is no single best model for all scenarios. New models can be incorporated as they are developed. A key issue for models is the trade-off between the quality of the embeddings and the computational overhead of the models. Producing the highest quality embeddings generally requires more overhead. It is generally preferable to use the highest quality model that can meet the response time requirements of the caching system. If the overhead for computing embeddings is too high, this will negatively affect cache throughput.

Even with the best models, semantic similarity will be inaccurate. Semantic similarity between two text strings, $s_1$ and $s_2$, is typically performed by converting $s_1$ and $s_2$ into vectors $v_1$ and $v_2$ respectively. The vectors are then compared using a similarity metric $S$ such as cosine similarity, Euclidean distance, dot product, etc. Let $t_s$ be a semantic similarity threshold. If 
\[S(v_1, v_2) > t_s\]
then $v_1$ and $v_2$ are considered to be similar. Otherwise, they are considered to be different.

Within a caching system, the semantic similarity threshold $t_s$ can be used to determine if a cache hit has occurred. If a user-issued query $q_1$ has a vector similarity with at least one other cached query that exceeds $t_s$, then a cache hit has occurred. The answer to the cached query with the highest vector similarity with $q_1$ is then used to satisfy the query.

This is a fairly rigid definition for semantic similarity, and it is only an approximate method. We argue that $t_s$ should vary based on multiple factors, including content of the query, run-time situations, and user preferences. Certain types of content should have a higher semantic similarity threshold. For example, LLMs can be used for generating computer code. Users need to give very precise instructions for asking LLMs to generate computer code in order to have a reasonable change of getting the LLMs to produce code which properly handles the task at hand. Requests for computer code should thus be associated with high similarity thresholds. By contrast, queries for natural language text can tolerate lower semantic similarity thresholds.

Another key aspect is that our system can assess a query in terms of the cost that would be incurred by sending the query to an LLM (both monetary and latency) and the likelihood of the request succeeding. Based on this assessment, the system can make a determination of how high $t_s$ should be. If a request is considered to have an elevated cost, then a lower value of $t_s$ can be used to increase the probability of a cache hit. By contrast, if a request is considered to have a lower cost, then a higher value of $t_s$ can be used to try to find a better match for the query.

The monetary cost associated with a request can be estimated in part by considering the LLM model. Some LLMs are free to access, while others charge varying amounts of money depending on the type of query. Latencies can vary considerably, from a small number of seconds to well over a minute. These costs can vary even from the same LLM provider. For example, OpenAI provides multiple models for its ChatGPT service. The different models have considerably different price rates. On May 13, 2024, the price for output from gpt-3.5-turbo-0125 was \$1.50 per million tokens. The price for output from the gpt-4-32k was \$120 per million tokens. In other words, gpt-4-32k output is 80 times more expensive than gpt-3.5-turbo-0125 output. For input, gpt-4-32k is 120 times more expensive than gpt-3.5-turbo-0125. Furthermore, the response times for gpt-4 models are generally considerably higher than the response times for the gpt-3.5 models. The gpt-4 models are more advanced than the gpt-3.5 models. However, there is a higher cost for the gpt-4 models. Thus, more money can be saved by avoiding requests to the gpt-4-32k model than by avoiding requests from the gpt-3.5-turbo-0125 model. It may thus be advantageous to use a higher value of $t_s$ for the gpt-3.5-turbo-0125 model than for the gpt-4-32k model.

Monetary cost is also a function of both query and response size. LLMs such as ChatGPT charge users based on tokens, and the API allows users to set a limit on the maximum number of tokens allowed in a response. Our system can consider both the query size and any user-specified token limit on response size to estimate the monetary cost of a request.

Expected response times are also an important factor to consider. If the expected response time for a request is high, this would suggest using a lower value of $t_s$. Response times are correlated with the LLM models, as well as with the size of responses. For ChatGPT, the more advanced models tend to have higher response times. For example, the gpt-4 models tend to have higher response times than the gpt-3.5-turbo models. Query and response sizes also affect response times. Larger query and response sizes are correlated with larger response times.

Lack of connectivity and unresponsive LLMs are other factors to consider. When connectivity to LLMs is poor, a higher percentage of requests should be satisfied from caches which are accessible. These situations warrant a lower value of $t_s$. There are also situations where the user has connectivity but an LLM is unresponsive. GenerativeCache allows users to be connected with multiple LLMs so that one LLM can compensate if another LLM is unresponsive.

User preferences should also be considered in determining values of $t_s$. This is particularly important when caching is being done on the client side. We provide programmatic and user interfaces allowing users to adjust caching parameters. This is particularly useful when users are personally interacting with LLMs, examining responses from the LLMs, and issuing new queries based on the responses. More details are contained in Section~\ref{section:enhanced}.

\section{Generative Caching}
A key aspect of GenerativeCache is generative caching. With generative caching, caches cooperate to combine multiple cached answers in response to user queries. This allows GenerativeCache to provide answers to queries which have not yet been received. In addition, GenerativeCache can provide new answers to queries which are richer than the individual responses which were originally received from an LLM.

As a simple example, suppose the system received the following queries: $Q_1$: “What is an application-level denial of service attack?” and $Q_2$: “What are the most effective techniques for defending against denial-of-service attacks?”. The answers to both of these queries are cached. Subsequently, the system receives a query $Q_3$: “What is an application-level denial of service attack, and what are the most effective techniques for defending against such attacks?” The system has not yet seen query $Q_3$, so the answer to $Q_3$ is not cached. Thus, using conventional caching techniques, a good answer to $Q_3$ is not stored in a cache because the answers to either $Q_1$ or $Q_2$ are incomplete.

Using generative caching, the answer to query $Q_3$ can be obtained by combining the answer to $Q_2$ with that of $Q_1$. Generative caching is thus able to satisfy query $Q_3$ using a combination of answers which have been cached.

Generative caching works by searching caches for more than one query which can satisfy a request query, $Q_5$. Responses are selected from among the cached queries which are closest in similarity to $Q_5$. Generative caching can be invoked in two modes: 
\begin{itemize}
    \item Primary: Generative caching is the default algorithm used to satisfy a cache miss.
    \item Secondary: Generative caching is only invoked if the regular cache lookup algorithm results in a cache miss.
\end{itemize}

Generative caching uses two parameters:
\[t_{single} < t_s\]
and
\[t_{combined} > t_s.\]

Let $S$ be a function which computes semantic similarity between two queries. Then our caching algorithm is:

\begin{algorithmic}
\State $X \gets \{$cached queries $x_i$ for which $S(x_i, Q_5) > t_{single}\}$
\If{$\sum_{x_i \in X} S(x_i, Q_5) > t_{combined}$} 
    \State cache hit has occurred
\Else
    \State cache miss has occurred
\EndIf 
\end{algorithmic}

In other words, we consider cached queries $x_i$ for which the semantic similarity between $Q_5$ and $x_i$ exceeds $t_{single}$. For all such cached queries $x_i$, the sum of the similarities of the queries with $Q_5$ has to exceed $t_{combined}$. If so, a cache hit has occurred. The system can either provide a combination of all answers obtained from the cache or perform a summarization of the answers and provide them to the user.

The answer generated for satisfying $Q_5$ can be cached for answering future queries semantically similar to $Q_5$.

\subsection{Optimizing Performance and Responses Based on User Feedback}

Users will often make multiple queries to a large language model before obtaining the exact answers they are looking for. In this scenario, it makes sense to try to provide at least some responses from the cache due to the low cost of doing so. Each query to a large language model can cost money and take several seconds to get a response. A response from the cache will take less than a second and will not incur a monetary cost.

Users can provide feedback on whether they are satisfied with a cached answer. Note that all cached answers provided to users are cache hits. If a user approves of a cache hit, it is known as a {\it high quality cache hit}. If a user is not happy with the cached response and obtains a better response from a large language model, it is known as a {\it low quality cache hit}. If a user is not happy with the cached response but the response from an LLM is not perceived as being better, than the response from the cache is still considered to be a high quality cache hit, as the cache cannot be expected to improve upon responses provided by LLMs.

When users are providing feedback, the system can keep track of the number of high and low quality cache hits. The quality rate is the number of high quality cache hits divided by the total number of cache hits. Users can set a target value for the quality rate, $t4$. Our system then varies the semantic similarity threshold parameter $t_s$ to achieve a quality rate close to $t4$. The value of $t_s$ is increased when the quality rate is below $t4$ by more than a threshold. The value of $t_s$ is decreased when the quality rate is above $t4$ by more than a threshold. Since the cost of a cache hit lookup is relatively small, the quality rate does not have to be very high. A lower quality hit rate will increase the cache hit rate, increasing the odds that more requests will be satisfied from the cache. Thus,

\begin{algorithmic}
\State $t4 \gets $ target value for quality rate
\State $quality\_rate \gets $ high quality cache hits / total cache hits
\If{$quality\_rate < t4$ by more than a threshold}
    \State increase $t_s$
\Else
        \If{$quality\_rate > t4$ by more than a threshold} 
        \State increase $t_s$
        \EndIf
\EndIf 
\end{algorithmic}

A key objective that our system tries to achieve is to minimize monetary costs for LLM queries. We can reduce monetary cost for an LLM such as ChatGPT using a number of methods, including:
\begin{itemize}
    \item Caching.
    \item Model selection. The more advanced models are generally more expensive to use. However, they can result in higher quality responses.
    \item Token limits for responses from LLMs. Lower token limits can reduce request costs. However, the responses will have less detail.
\end{itemize}

A key point is that we can adjust the cache hit rate to control the monetary cost of making request to an LLM. For ChatGPT, the monetary cost is a function of both the model used and the token limit. However, different LLMs can have different methods for charging money for requests. The cache can be adjusted to reduce LLM costs based on a target cost. The user can set a preferred cost per request, c. As our system makes requests, it monitors them and calculates the cost from the number of tokens in the request and responses, as well as the model. If the cost per request is too high, it reduces the cost per request by caching. The key point is that we can control the cache hit rate by adjusting the semantic similarity threshold. If the preferred average cost per request is c1 and the average cost per request with no caching is c2, where c2 > c1, we can adjust $t_s$ so that the cache hit rate approaches (c2 – c1) / c2.

Another aspect is that if we have feedback from users, we can tune the system to make fewer calls to the LLM.  The idea is to adjust $t_s$, to serve as much content from the cache as possible while still providing cached answers the user is satisfied with. The system lowers $t_s$ as long as the user is satisfied with the cached responses. When the user expresses dissatisfaction with the quality of the cached responses, $t_s$ is increased. This process results in a $t_s$ value which maximizes the cache hit ratio while providing good responses to the user.

Costs to use an LLM can further be reduced by tuning the request parameters associated with the LLM. For ChatGPT, different models can be experimented with to judge the satisfaction level of users. For example, the GPT-4 models are generally more powerful but also more expensive than the  GPT-3.5 models. If users are flexible regarding choice of models, then GenerativeCache will provide some responses using the cheaper, less powerful models. If the users are satisfied with the responses, requests will continue to be served from the less powerful models. Otherwise, requests will be served using more powerful models.

Token limits can also be used to limit costs. If the user is fine with shorter responses, GenerativeCache saves money by using lower token limits.

\section{Caching Architecture}
Caches can be deployed at multiple points within a network. These locations include at the client, within the network, as well as on the server. In this section, we discuss different architectural options for caches. 

We have different caching architectures for different scenarios. In the case of a cache for a single client, our cache can be integrated as a single executing program within the client. The cache can function as a single process. There is thus no need to have multiple processes and interprocess communication. We have developed a Python-based cache in which Python-based APIs are used to communicate with LLMs. For faster performance, cached data is maintained in memory. For fault tolerance, the contents of a cache can be stored persistently.  That way, if the caching process fails, the contents of the cache can be restored from disk. Our system also allows a cache to be brought to a warm state quickly. The cache can be loaded with query-answer pairs from past sessions. That way, it is possible to maintain cached query-response pairs over time and load the most relevant query-response pairs to accommodate different sessions.

A cache which is serving multiple clients needs to have a higher throughput than one which is serving a single user. For these scenarios, we make use of Redis~\cite{carlson2013redis}. Redis provides a scalable key-value store. However, the key-value capabilities provided by Redis are not sufficient for our purposes. Our system performs lookups using semantic similarity. We thus need efficient lookups using vectors. There are a number of vector databases~\cite{pan2023survey} that offer high performance. A key aspect of vector comparisons is that there is considerable overhead for making exact comparisons. For traditional caches, key-value stores are sufficient. They are typically implemented as a hash table, for which the average lookup time is constant. By contrast, finding the closest match to a vector is not as straightforward. Simple methods can have high overhead due to the amount of searching that is needed.

There are approximate methods which can identify similar vectors efficiently.  A wide variety of vector databases are available such as Milvus, Weaviate, Pinecone, Chroma, and others. These vector databases typically have scalable performance. We have used Milvus~\cite{wang2021milvus}. While the overall performance of Milvus is good and Milvus is quite effective in terms of its vector handling capabilities, it has limited capabilities for storing data other than vectors. Milvus has a VARCHAR field for character strings of up to 65,535 characters. This limitation can be a problem. For data which are not text (e.g. binary data), the data would have to be serialized. For data which are too big to fit within this field, it would be necessary to use a separate database in addition to Milvus. Milvus would store vectors and a key to reference the data object in another database. Having multiple databases introduces complexity. It would be easier to just require a single database. Furthermore, there is some additional overhead required due to the fact that two database lookups are required. The first lookup would be to Milvus to find the closest match vector (s). The second lookup would be to the other database to find the actual data corresponding to the vector (s) identified by Milvus.

Ideally, we would prefer to use a single database for both vectors and the data associated with the vectors. Redis Stack provides this capability. Redis is well known for providing a key-value store which can be distributed across multiple nodes. In this mode, it can function as a high-performance cache. Other features have also been added over the years, including RediSearch, which provides querying, indexing, and full-text search. RediSearch provides vector searching capabilities. RediSearch is currently available as part of Redis Stack which is open source and bundles together RedisJSON, RedisSearch, RedisTimeSeries, and RedisBloom (for Bloom and Cuckoo filters).

Redis Stack is sufficient for both efficiently managing vectors and storing the actual data. There is no reason to introduce another database. This makes it an attractive option, along with its good performance. It should be noted that the programmatic interface for vector queries has changed since the time before RediSearch was bundled with Redis Stack.

Our architecture allows different storage solutions to be bundled together for handling caching. For example, Milvus can be used instead of Redis Stack. For this option, another database should be used in conjunction with Milvus unless each query result being cached would fit within 65,535 characters.

The features offered by these data stores is constantly evolving. For example, Milvus did not offer a field for character strings before Milvus 2.1 which was released in 2022. As the software evolves, different data store options may become more preferable. For example, one might prefer to use a certain set of vector comparison algorithms used in a particular piece of software. In that case, our system would allow the software to be used for data storage as long as the appropriate APIs are implemented for managing and querying data within the data store.

Another option is to use libraries for calculating vector similarity such as Faiss~\cite{douze2024faiss} and deploy them for very specific vector similarity algorithms. That allows one to use customized vector similarity algorithms and not an algorithm tied to a particular vector database. A complication with this approach is that the data typically needs to be stored both in memory and on disk. Disk storage is important to prevent loss of data in the event of a process failure. It is also important if the size of the data exceeds main memory. A proper implementation thus involves managing what is stored in main memory as well as on disk. Because of complications such as this, it is not entirely straightforward to produce a high-performing implementation using a customized vector comparison algorithm. Using an established vector database avoids these complications and would be a considerably more straightforward approach.

In the most general case, our system utilizes multiple caches which can cooperate and share content. A level 1 (L1) cache would be associated with a particular client. One or more level 2 (L2) caches would be shared by several clients. An initial request from a client would first try to obtain the data from an L1 cache. In the event that the L1 cache cannot satisfy the request, the system tries to obtain a response from an L2 cache.

In order to handle a large number of clients, there could be multiple L2 caches. Each L2 cache functions as an L2 cache for a different set of clients. The L2 caches communicate to share content. When a response cannot be found in one cache, another cache can be contacted to provide a response.

\begin{figure}
  \centering
  \includegraphics[width=\linewidth]{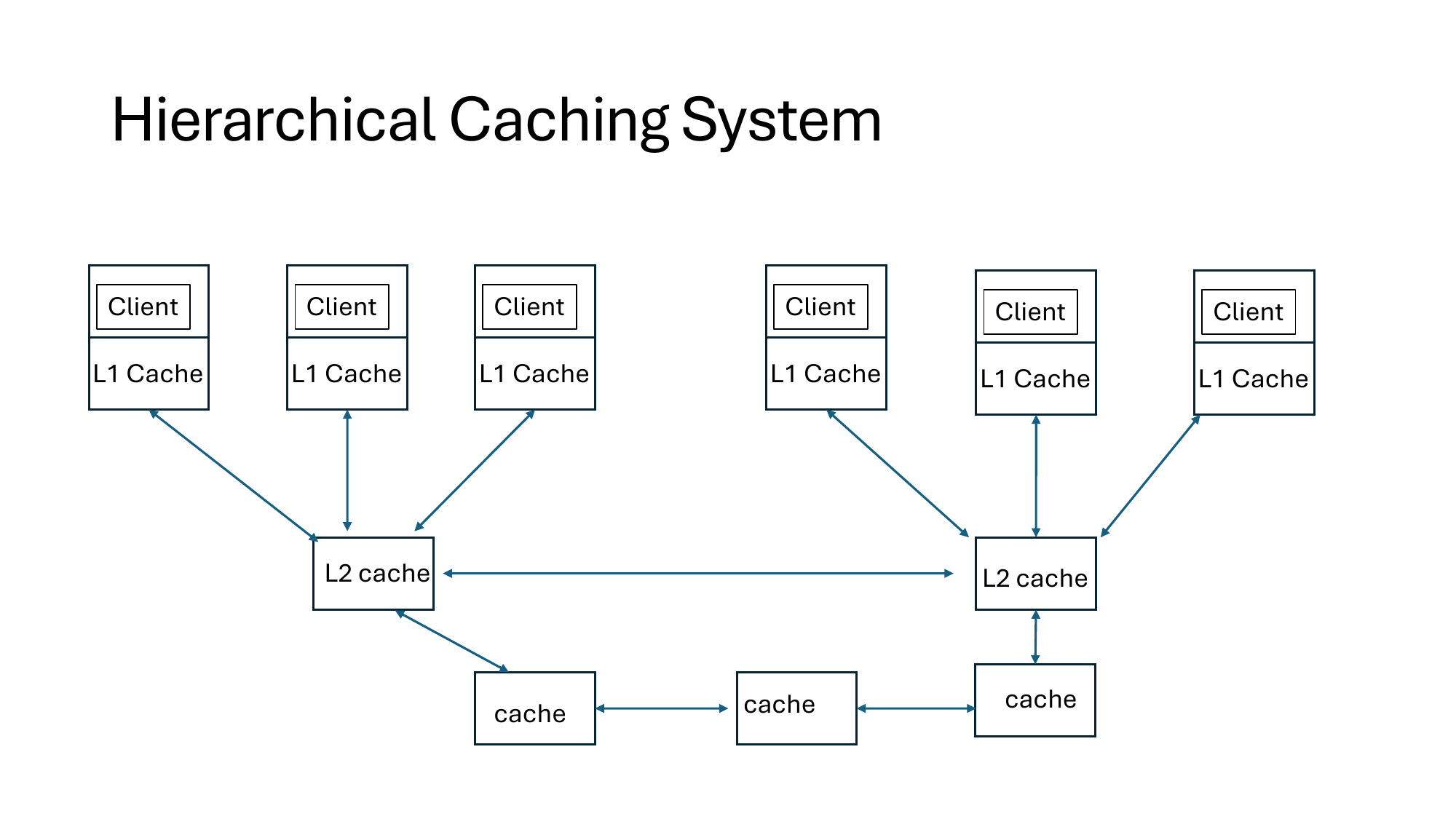}
  \caption{A distributed hierarchical generative caching system.}
  \label{fig:distributed}
\end{figure}

Figure~\ref{fig:distributed} shows how GenerativeCache is deployed in a hierarchical distributed caching system. Clients have a local L1 cache. When a request is made by a client, the request first goes to the client L1 cache. The similarity threshold, $t_s(1)$, depends on the factors described earlier. In the event of a cache miss, the request goes to the L2 cache corresponding to the client; $t_s(1)$ is used as the similarity threshold for this cache as well. If the L2 cache is able to satisfy the request with a query-response pair q1, q1 is then stored in the L1 cache. Otherwise,  the L2 cache can contact other caches. The other caches also use $t_s(1)$.

In the event of a cache miss in either the L1 or L2 cache, the query-response pair used to satisfy the request is stored in the cache.

Some queries might be more personalized than others. In other words, an individual  may be making queries to an LLM which are specific to that individual. However, those requests may not be of significant interest to the wider population. In that situation, it may be advisable to cache the content within an L1 cache specific to the client. However, it would be less appropriate to cache the queries in an L2 cache shared by a wider community of users. 

The system could classify the types of queries to determine which could be cached based on the nature of the queries. However, this gets complicated. Instead, we give clients some control over whether the response from an LLM should be cached. Users can indicate that a particular response should not be cached at either the L1 cache, the L2 cache, or both. This allows users to provide some hints to facilitate more effective caching. It also allows users to keep query-result pairs which they consider too personal from being cached.

The system can also be configured to follow certain policies when multiple caches are present. For example, in situations in which there is an L1 and L2 cache, it is possible to have all query-response pairs which are in the L1 cache also be present in the L2 cache (inclusion). In the general case, though, some but not all of the data in L1 cache will also be in the L2 cache.

Multiple caches can also contribute in providing generative content. Generative caching synthesizes responses to a query from multiple cached query results. Having a larger pool of content from several caches allows a richer set of responses to queries than just using a single cache.

Contacting more caches to provide generative content adds overhead. The system can thus limit the processing and communication to provide cached responses with lower overhead. It should be noted that the relatively high cost of contacting LLMs for requests means that even if the caching system takes a bit of extra time to provide a response, the performance and cost savings provided by the cache can still be quite significant. This means that an LLM caching system such as ours has more leeway to spend a bit longer to generate a really good response than caches in other distributed environments, such as web proxy and CDN caches.

\section{Enhanced Client for GenerativeCache}
\label{section:enhanced}
In order to most effectively use LLMs, good client software is essential. In this section, we describe the enhanced client software for GenerativeCache. A key aspect is the way in which the client software is integrated with the caching system to enhance both performance and quality of responses returned to the user.

\subsection{Different Ways of Interacting with Large Language Models}
There are multiple ways to interact with LLMs, and the different modes of interaction affect what type of caching policies will be most effective. One option is for users to interact directly with the LLM via a browser or other user interface, issue queries, interpret the responses, and make additional queries based on responses from the LLM. We refer to this mode of communicating with a large language model as {\it interactive mode}. We have designed a UI specifically for interactive mode. The user can issue queries to multiple LLMs with specific parameters. The user can also specify different semantic similarity thresholds ($t_s$) and other parameters. Our system provides performance statistics from the LLMs. Users can adjust $t_s$ for caches based on their experiences. Lower  $t_s$ values allow higher hit rates. Higher  $t_s$ values result in more requests to the LLMs.

Caches can also operate in a mode which is more independent from human intervention. An LLM cache can be placed within the network and function in a manner similar to a proxy or CDN cache. We refer to this mode as {\it automatic mode}. Caches in automatic mode are not explicitly managed by users. Instead, they are configured to react to run-time conditions and function based on these run-time conditions. The cache can be configured to adapt to different run-time conditions in automatic mode. For example, it can estimate expected costs of the requests, as well as maintain information on response times from different LLMs. Based on this information, it can adjust  $t_s$ appropriately. 

Many of the features for the GenerativeCache enhanced client are designed for interactive mode.

\subsection{Key Features of the Enhanced Client}

The enhanced client for GenerativeCache provides the capability to make queries to multiple LLMs. The LLMs provide a programmatic API for using them.  The enhanced client incorporates Python APIs for communicating with LLMs such as ChatGPT~\cite{van2023chatgpt}, Gemini~\cite{team2023gemini}, and Llama 2~\cite{touvron2023llama}. GenerativeCache caches responses from the various LLMs, depending on how caching parameters have been set. 

Multiple responses may be cached for the same query. This could be a result of situations such as the following:
\begin{itemize}
    \item The query was answered by multiple LLMs.
    \item The same query was issued with multiple sets of parameters.
    \item A similar query was already cached with response $r_1$. However, the user explicitly requested a new response from an LLM and not $r_1$. After the new response $r_2$ is received from an LLM, $r_2$ is added to the cache, along with the previously cached $r_1$. 
\end{itemize}

In order to enable the user to easily make a variety of requests to the LLMs and try out different queries and parameter settings, we provide both a programmatic API in Python as well as a graphical user interface for allowing users to query the various LLMs and make use of generative caching. A REST (HTTP) interface is another option.

Because of the significant response times that remote calls to a large language model might incur, making multiple calls to LLMs can consume several minutes. One such scenario would be making the same query to multiple LLMs while trying out various parameters. In order to reduce this waiting time, it is essential to have parallelism in which multiple calls to LLMs can be made concurrently.

We provide both synchronous and asynchronous methods of invoking LLMs. The asynchronous approach allows multiple requests to be made in parallel. These asynchronous methods are important for getting decent performance when multiple requests are being made and the response from one request is not required before invoking another request.

The asynchronous interfaces are implemented in Python using the asyncio library. This design works because LLMs like ChatGPT and Gemini provide Python API calls which issue requests using non-blocking IO.  The asynchronous interfaces have low overhead. Therefore, the speedup which is achieved for calls to ChatGPT and Gemini is close to linear. 

If the Python interface for accessing a large language model does not provide a function for accessing the LLM using non-blocking IO, then multiprocessing is used to generate parallel requests. The asyncio library will be ineffective in this situation. Furthermore, the global interpreter lock (GIL) would limit concurrency. Therefore, multiple processes are needed for concurrency.

\begin{figure}
  \centering
  \includegraphics[width=\linewidth]{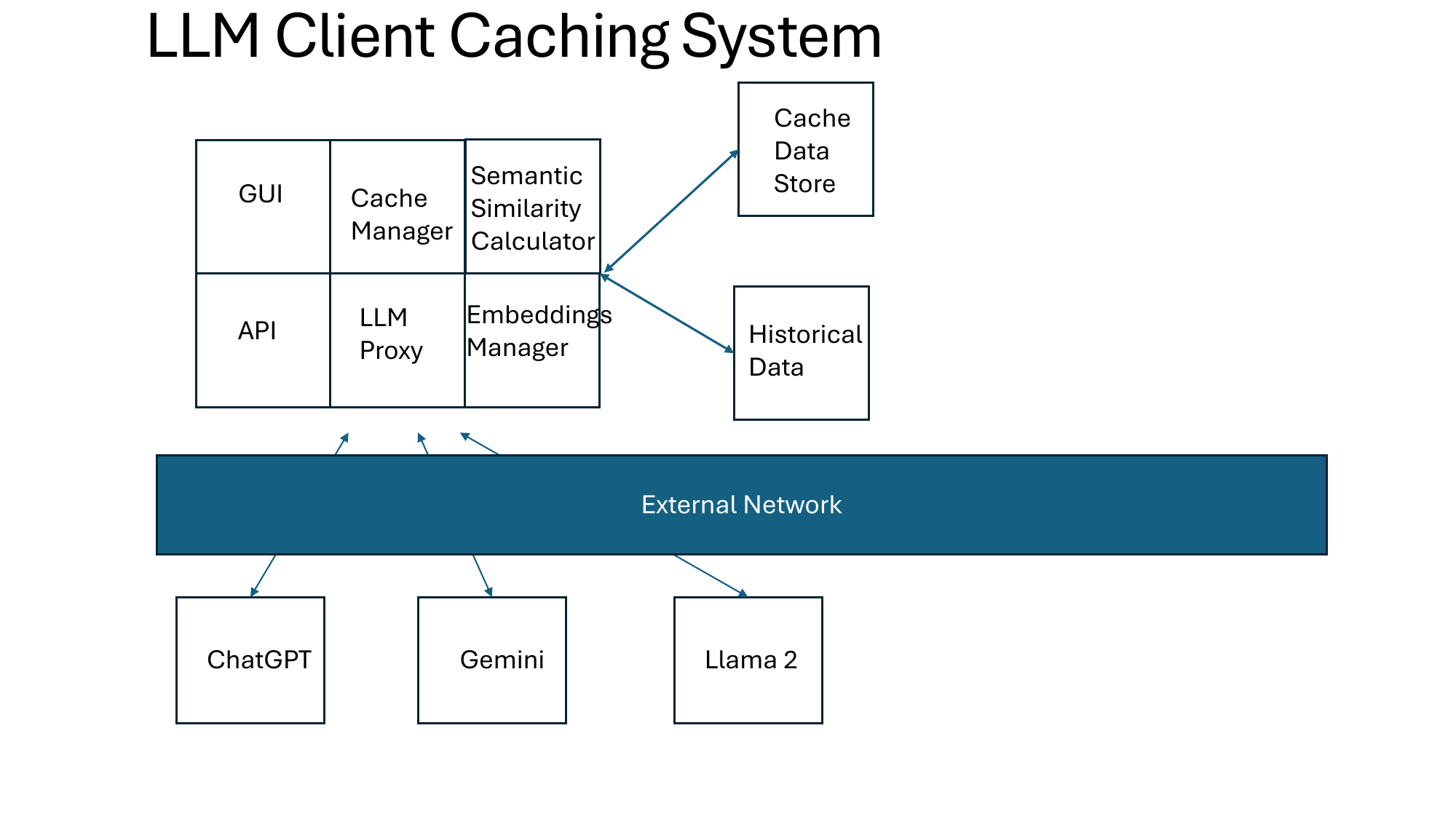}
  \caption{Our enhanced client for LLMs, with integrated caching.}
  \label{fig:overall}
\end{figure}

Figure~\ref{fig:overall} depicts the enhanced client for GenerativeCache. 
The cache manager is the software that manages the key operations of the cache and implements the caching algorithms. The LLM proxy is the software that manages interactions with LLMs, makes requests to them, and combines the output from the various LLMs. The LLM proxy provides both sequential and parallel interfaces for querying LLMs.

Determining semantic similarity of queries is an essential component. This is provided by the semantic similarity calculator. In some cases, semantic similarity will be provided by a vector database. In other cases, more customized algorithms can be used for calculating vector similarity. 

The embeddings manager embeds natural language text within vectors. The exact method for doing so do depends on the model. A wide variety of models can be used by our system. It is expected that new models will continuously be plugged in and used with our system. 

We provide both a graphical user interface (GUI) as well as a programmatic API in which our system can be used as a Python library and invoked by Python functions. The GUI is appropriate for using our system in interactive mode. The programmatic API is appropriate for using our system in automatic mode.

\subsection{A Modular Architecture Supports Multiple Pluggable Components}

The software architecture of GenerativeCache is a modular one in which different data stores can be used for storing the data, and different methods can be used for computing embeddings of queries in order to determine semantic similarities of the queries. 
The methods for computing embeddings depends on the model that is used. The methods for computing embeddings are responsible for obtaining the appropriate models and using them to calculate semantic similarities of queries.

\begin{figure}
  \centering
  \includegraphics[width=\linewidth]{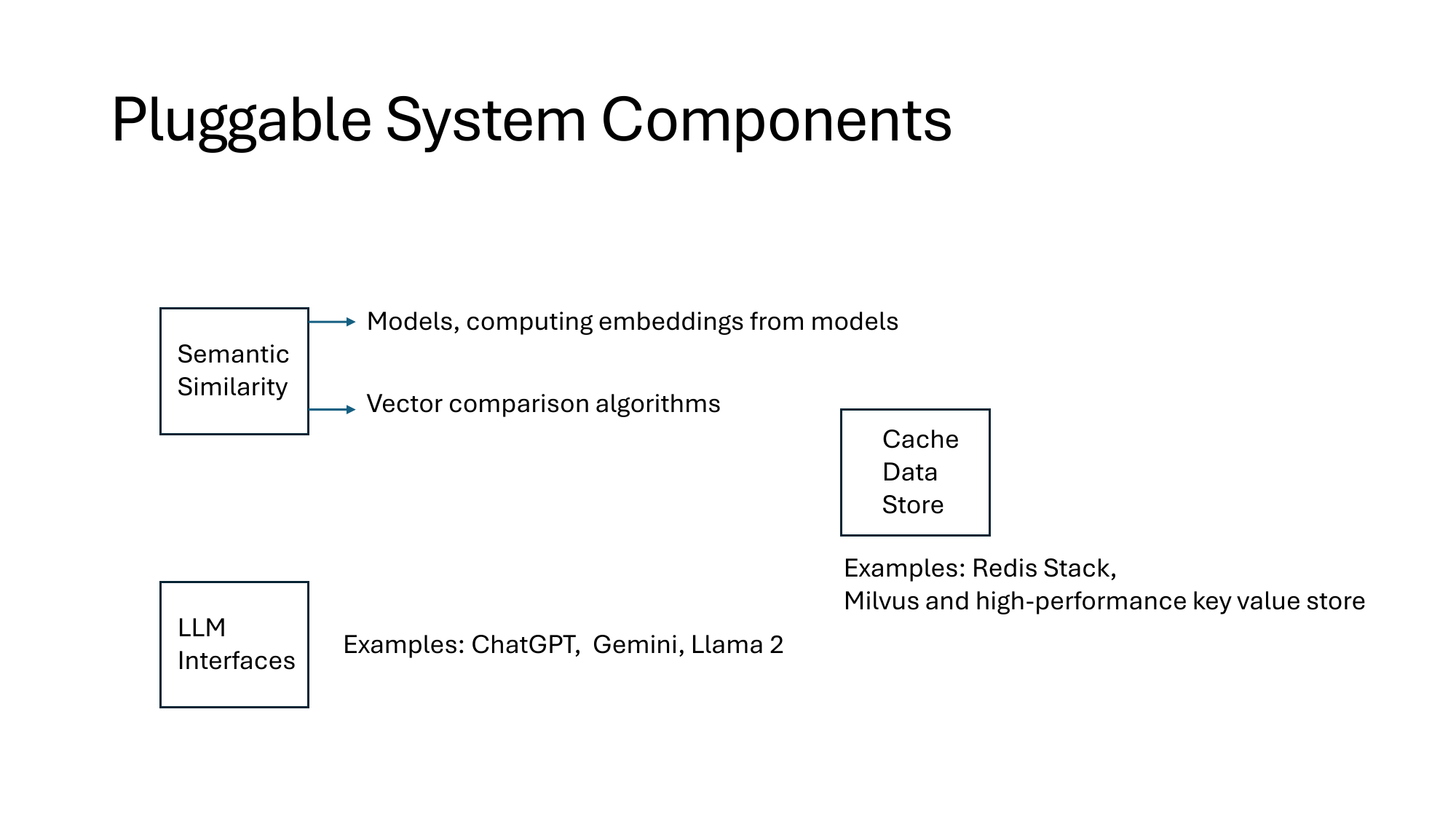}
  \caption{Key components of our system which can be customized and swapped out for other components.}
  \label{fig:components}
\end{figure}

Figure~\ref{fig:components} depicts software components of our system which are intended to be customized for specific use cases. Default options exist, with a wide variety of other possibilities. The key components of the system which are intended to be customized include:
\begin{itemize}
    \item The semantic similarity calculator.
    \item The data store (s) used for the cache.
    \item The API functions for communicating with different LLMs.
\end{itemize}

The semantic similarity calculator needs to embed text within vectors and compute vector similarity. Models for calculating text embeddings are constantly evolving. It is thus expected that users would want to try a wide variety of different models. Depending on which models are used, the actual code for determining embeddings can vary. In order to use different models, certain functions for calculating embeddings using the models need to be provided.

In order to select models, benchmarks like the Massive Text Embedding Benchmark (MTEB)~\cite{muennighoff2022mteb} can be used for comparing different models. Models are constantly changing and being improved. This is a key reason why it is important to allow different models to be plugged into our caching system. As models change and improve, new ones should be substituted for the older ones. The MTEB leaderboard ranks models and can be used for identifying the current best ones.

It should be noted that while such rankings are helpful for comparing models, they do not tell the whole story. Certain models are better than some tasks than others. Monetary cost may be a consideration, as some models are free while others are proprietary and cost money. It should also be noted that not all of the models work as claimed. Because there are so many models available, it is often prudent to move onto a new model if the model that one first chooses does not work.

Another key aspect of determining semantic similarity is calculating the similarity of vectors corresponding to embedded text. In some cases, this can be done using a vector database. Another option is to provide customized methods of doing comparisons without resorting to a vector database. We have used both of these approaches. The vector database approach is appropriate for a scalable implementation which can handle high request rates using proven technology. The customized method is for limiting dependencies on external databases. This version of our generative caching system can run as a Python library within a single process. It is a lighter weight version with fewer dependencies that is easier to use but will not have the same degree of scalability as the vector database versions. It is appropriate for a single client, but would be less suitable for handling requests for multiple clients.

The data store used for storing cached data is a critical part of the system. Our architecture allows multiple data stores to be used. One solution which has worked well for us is Redis Stack. This functions as both a vector database as well as a store for the data. Another option that we have used for storing vectors is Milvus. While Milvus has good scalable performance, its capabilities for storing the actual data are limited. It will generally be necessary to use another data store for the data itself, with Milvus storing the key for the data along with the vector. This means that two data stores will have to be maintained. Fetching data from the cache, as well as storing data in the cache, will consume two sequential database operations.

The use of Redis Stack and Milvus are appropriate for caches accessed by several users wherein the request rates can be substantial. For situations with a cache accessed by a single user, it is possible to use a lighter weight solution which does not rely on databases like Redis Stack or Milvus. We have implemented an in-memory data store option wherein all of the cached data and vector computations are performed in memory. The contents of the cache, along with the vector data, can be periodically backed up on disk so that it is available if the cache process fails or is intentionally terminated. This is a lighter weight option which is feasible for a single client. It does not have the scalability of the Redis Stack or Milvus options.

Our system is designed to be compatible with many different LLMs. The LLM interfaces depicted in Figure 2 are thus a critical part of our system. LLMs such as ChatGPT and Gemini provide Python libraries allowing Python programs to access them. It is also possible to send requests to these LLMs via REST (HTTP).

Different LLMs offer different parameters within their APIs. For example, the ChatGPT API allows users to select different roles, models, temperaturess (which controls the degree of creativity or randomness in the output), as well as optional limits on the maximum number of tokens in the response (max\_tokens). The model selected, as well as the max\_tokens parameter, can have  a significant effect on monetary cost as well as response times.  Our system makes use of these parameters to optimize the quality of responses while limiting monetary cost and response times.


\section{Experimental Results}

The major overhead for our caching system is calculating embeddings for queries. Cache lookups and additions consume considerably fewer CPU cycles. 

We have tested our cache using the Stanford Question Answering Dataset (SQuAD)~\cite{rajpurkar2016squad}. This data set contains over 130,000 questions and answers. Performance numbers were collected when the cache executed on a 24 GHz Intel® Core™ i7-12800H Processor running Ubuntu 22.04.3 LTS.

\begin{figure}
  \centering
  \includegraphics[width=\linewidth]{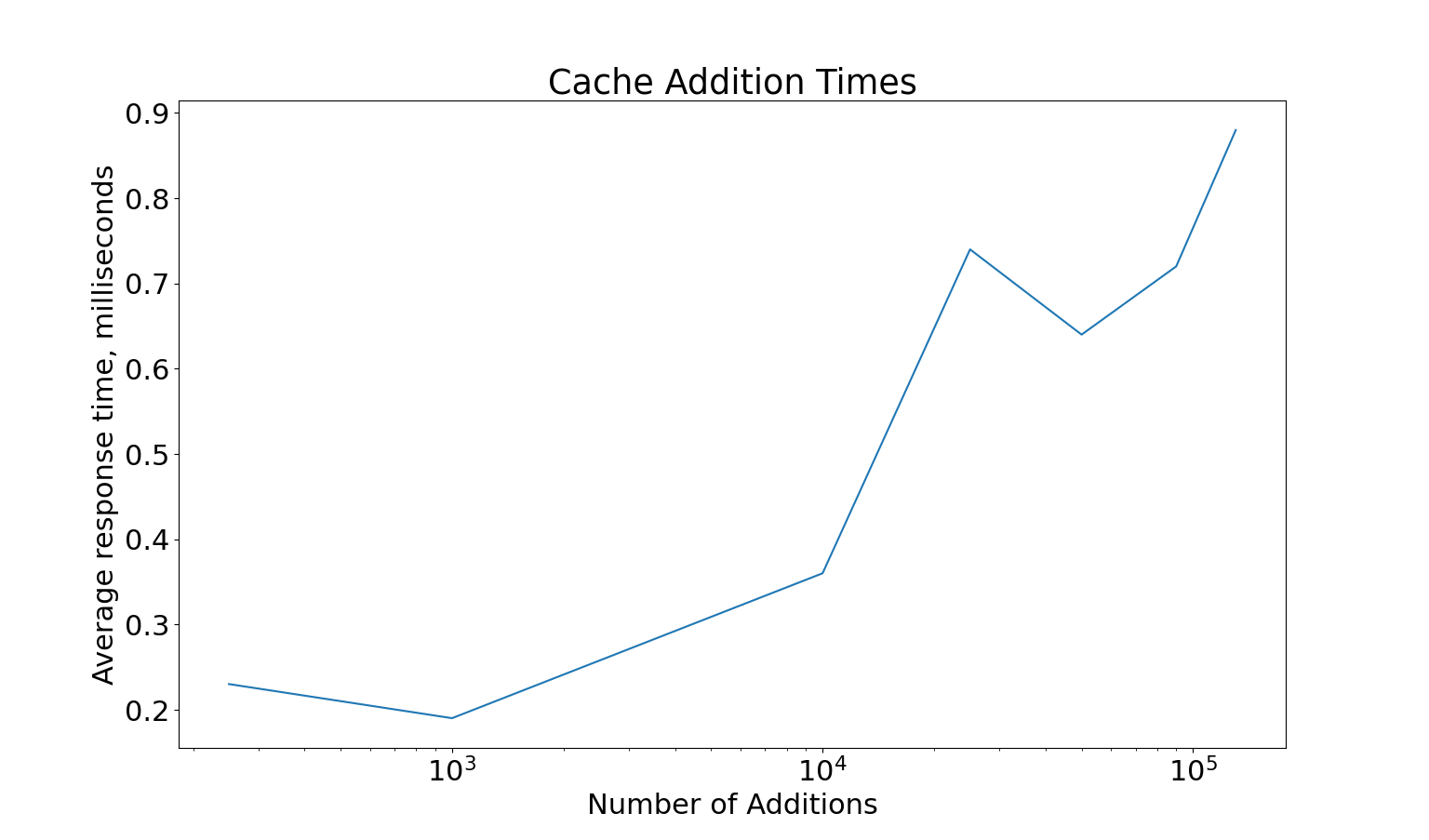}
  \caption{Average time in milliseconds to add query-result pairs to a cache.}
  \label{fig:graph-add}
\end{figure}

Figure~\ref{fig:graph-add} shows the average time in milliseconds to add a question-answer pair to a cache. Experiments started from an empty cache. The X-axis values represent the number of question-answer pairs added to the cache in each experiment. The overhead grows as more question-answering pairs are added to the cache.

\begin{figure}
  \centering
  \includegraphics[width=\linewidth]{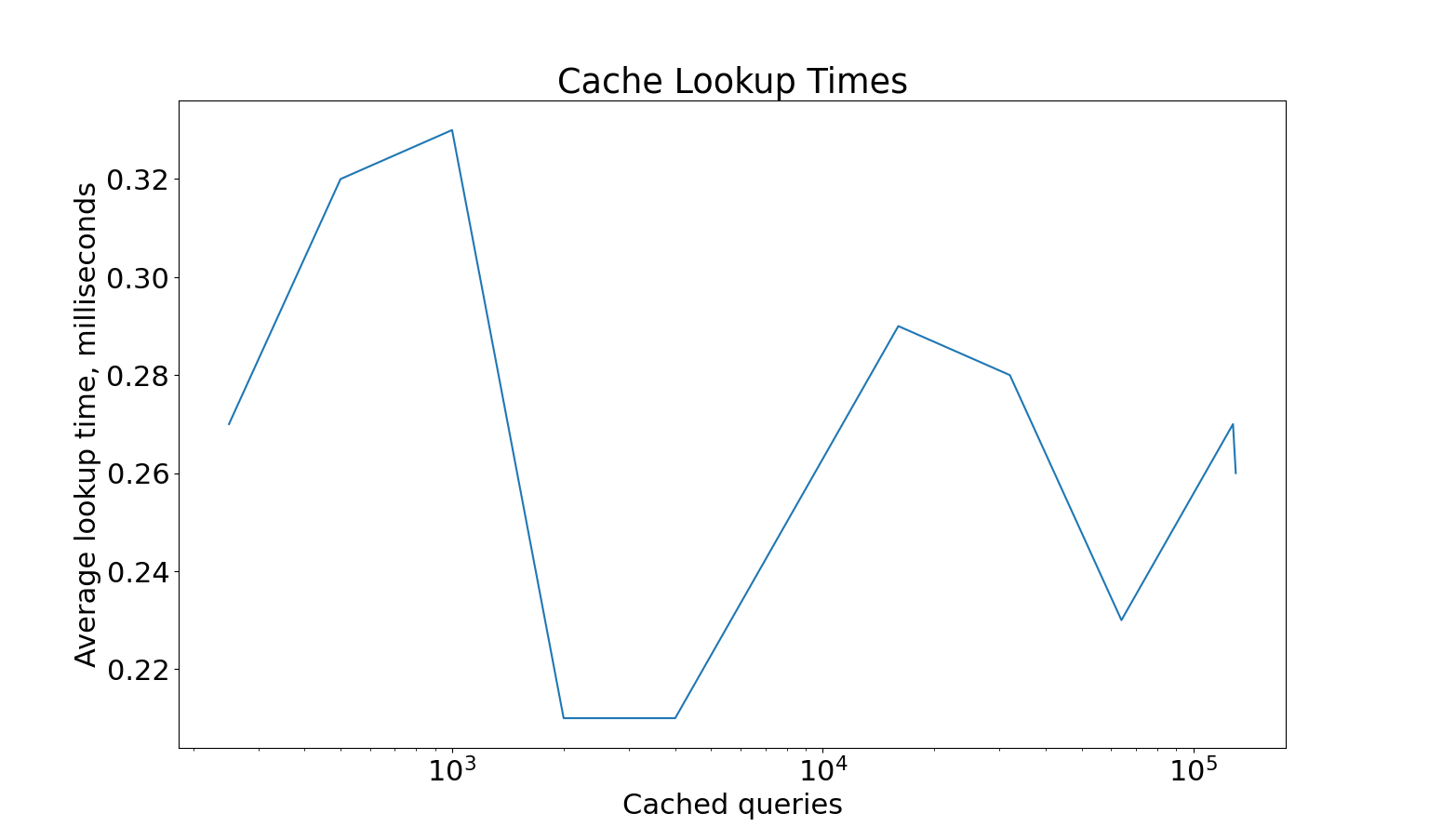}
  \caption{Average time in milliseconds to lookup query-result pairs in a cache.}
  \label{fig:graph-lookup}
\end{figure}

Figure~\ref{fig:graph-lookup} shows the average time to perform a cache lookup. The overhead does not grow with the number of cached question-answer pairs.

\begin{figure}
  \centering
  \includegraphics[width=\linewidth]{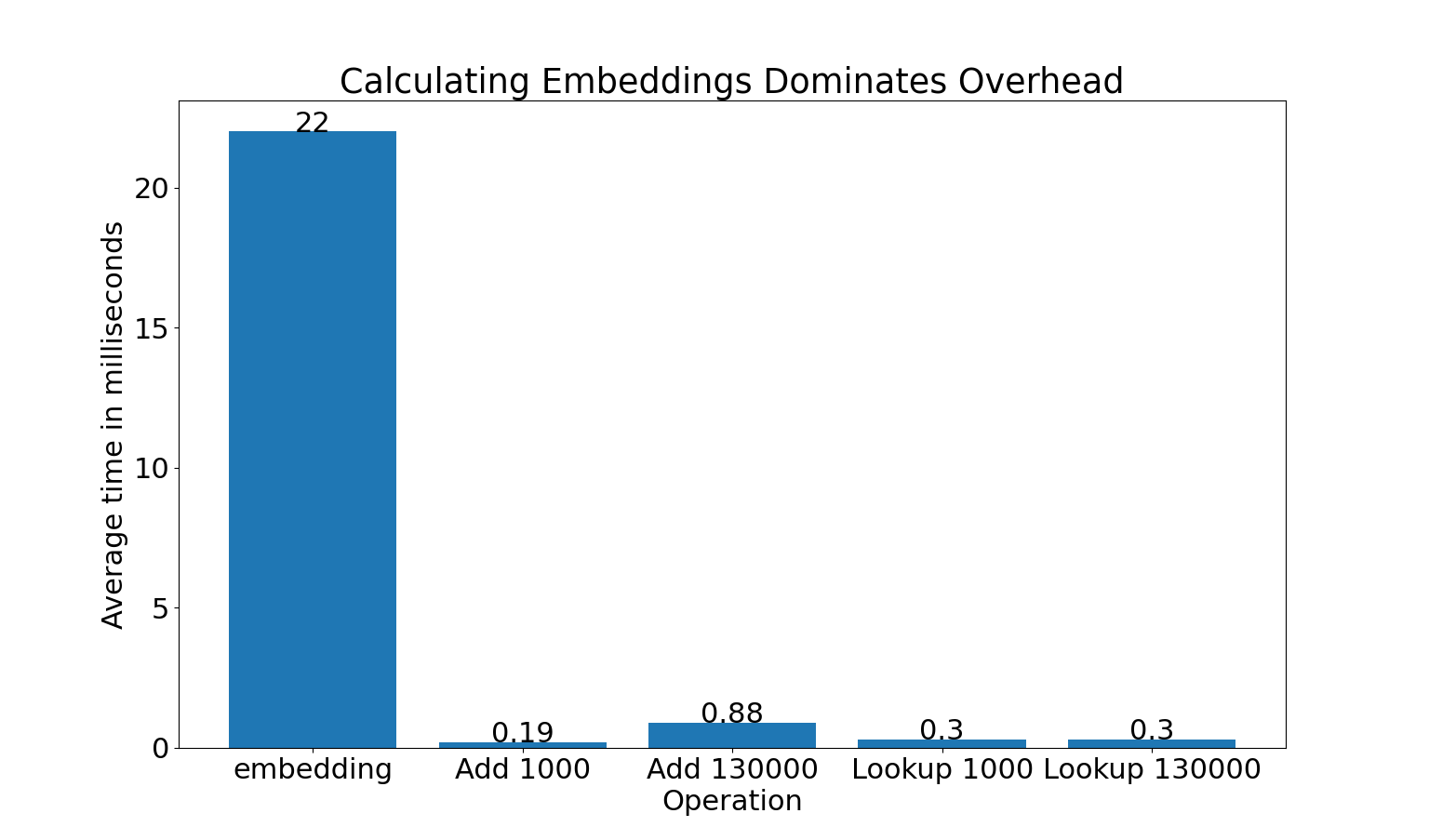}
  \caption{Overhead is dominated by the time to compute embeddings for queries.}
  \label{fig:graph-embeddings}
\end{figure}

Figure~\ref{fig:graph-embeddings} compares the overheads for various operations related to caching. The overhead is dominated by the time for computing embeddings. The first bar represents the average time to compute an embedding for a query. This overhead is the major one for the cache (22 ms). Facebook's msmarco contriever model was used to generate embeddings~\cite{embedding_model}. 

Without resorting to parallelism or the use of GPUs, GPTcache can sustain a throughput of about 45 requests per second on the hardware test system described above; as Figure~\ref{fig:graph-embeddings} shows, this overhead is dominated by the time to compute embeddings. The computational time can be reduced using high-performance GPUs. If many embeddings are being computed, parallelism can be used to speed up calculations.

The second and third bars are average times for adding a question-answer pair to the cache. In the second bar, 1000 question-answer pairs were added to an empty cache. In the third bar, 130,000 question-answer pairs were added to an empty cache. The average time to add to the cache grows with the number of cached query-answer pairs. The fourth and fifth bars are average times for performing a cache lookup. In the fourth bar, the cache contained 1000 question-answer pairs. In the fifth bar, the cache contained 130,000 question-answer pairs. Average lookup times did not increase as cached question-answer pairs increased within this range.

\begin{figure}
  \centering
  \includegraphics[width=\linewidth]{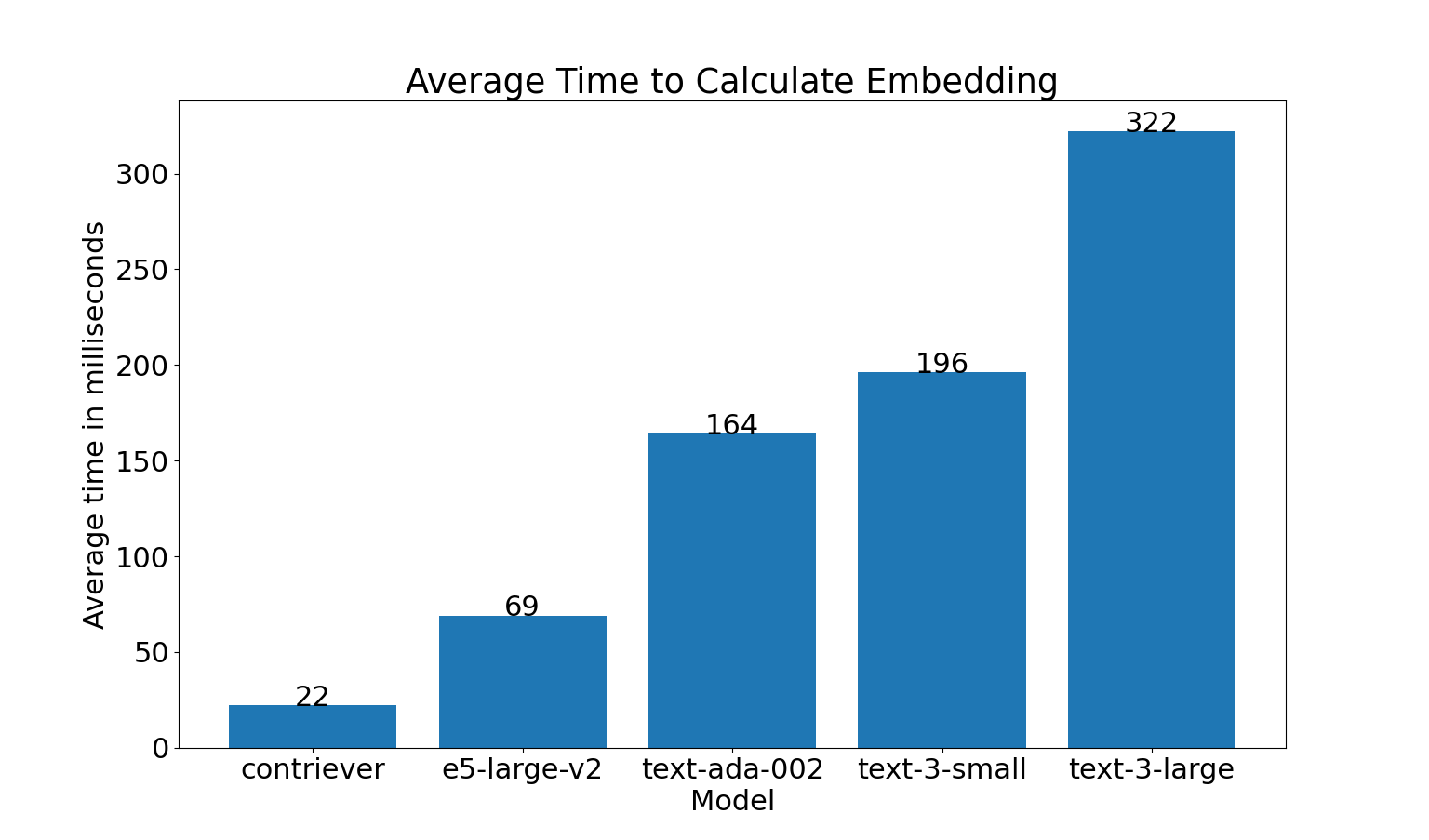}
  \caption{Average time for computing embeddings for five commonly used models.}
  \label{fig:graph-model-overhead}
\end{figure}

The time for computing embeddings varies considerably based on the models used. Figure~\ref{fig:graph-model-overhead} shows the average time to compute an embedding for the SQUAD questions using five commonly used models. The first bar corresponds to Facebook's msmarco contriever model~\cite{embedding_model}. The second bar corresponds to the e5-large-v2 model~\cite{embedding_model2}. The remaining bars correspond to the three models OpenAI is currently promoting for embeddings, namely text-embedding-3-small (fourth bar), text-embedding-3-large (fifth bar), and text-embedding-ada-002 (third bar). The msmarco contriever model is the fastest followed by the e5-large-v2 model. Both of them run locally on our test system and are free. By contrast, the OpenAI models run remotely on OpenAI's servers and cost money to use. The text-embedding-3-large model is currently OpenAI's most powerful embedding model but is also the most expensive to use. It also has the highest overhead in Figure~\ref{fig:graph-model-overhead}. The text-embedding-3-small model is more powerful and cheaper than the text-embedding-ada-002 model; however, the bargraph shows that the text-embedding-ada-002 model has lower latency.

The high latency as well as the cost of the remote OpenAI models are a drawback. There are high performing models which can run locally with considerably lower latency and cost. The use of GPUs and parallel processing can further increase performance and scalability of locally running models.

\subsection{Comparison with GPTcache}
GPTCache~\cite{bang2023gptcache} is an open source semantic cache designed for LLM content which is probably the most widely known one. It does not provide generative caching or the ability to tune and customize the semantic similarity algorithm to different workloads as we do. We have also found its performance to be considerably below GenerativeCache. We are able to sustain a throughput of 5 cache lookups per second. The average response time was .2 seconds per request. GenerativeCache thus is about 9 times faster than GPTCache.

The GPTCache paper, as well as the github repository, do not adequately performance and scalability. The default recommended configuration uses SQLite for cache storage. This is a poor choice for both semantic caches and conventional key-value caches because the cache queries are not inherently relational. Relational queries incur significant overhead which can only be justified if the inherent nature of the queries is relational. 


\section{Related Work}

GPTCache~\cite{bang2023gptcache} is an open source semantic cache for LLM content. GPTCache allows users to customize the cache for different embedding models, similarity assessment, and eviction policies. Portkey~\cite{Portkey} offers semantic caching for LLMs. They claim to get 99\% accuracy with an approximately 20\% hit rate. For retrieval-augmented generation, they claim hit rates range from 18-60\%. MeanCache~\cite{gill2024privacy} is a semantic caching system that places a local cache in each user’s device to preserve privacy. Federated learning is used to collaboratively train a query similarity modle across LLM users while preserving privacy. Regarding potential savings from caching, the authors find that 31\% of a user’s queries are similar to at least one previous query by the same user. 

None of these past works use generative caching. Furthermore, none of them coordinate the caching algorithms with the queries to achieve an optimal balance of reduced cost and quality of data as we do.

Cache replacement for LLM caches is studied in~\cite{zhu2023optimal}. The problem of whether a pair of queries can be satisfied by the same response is studied in~\cite{zhu2024efficient}. The authors tune embeddings for this purpose. Testing semantic caches is addressed in~\cite{rasool2024llms}. The authors generate test input for semantic caches by using LLMs.

LLM caching is based on past work that has been done in web caching~\cite{wang1999survey, Iyengar97c, Challenger99a} and query caching. Past work in caching search engine queries provides additional insight into cache hit rates which may be achievable. Baeza-Yates et al.~\cite{baeza2007impact} and Cambazoglu et al.~\cite{cambazoglu2010refreshing} claim that hit rates for cached search engine queries can approach 50\%. This improves upon previous works which found hit rates ranging from 29-43\%~\cite{ markatos2001caching, xie2002locality}.

Memcached~\cite{Memcached, Nishtala13}  and Redis~\cite{carlson2013redis} are open-source caches which have been widely used for improving performance in many web, cloud, and distributed environments. While earlier versions were primarily key-value stores, more recent versions of Redis support semantic caching. Vector database capabilities are offered by Redis Stack. The LangChain framework~\cite{topsakal2023creating} for developing LLM-powered apps provides a semantic cache that uses Redis as a vector-store backend. 

\section{Conclusion}

We have presented GenerativeCache, a generative caching system for large language models. The generative caching system allows caches to synthesize responses to new queries which have not yet been seen from existing cached responses. We improve upon previous semantic caching methods by adaptively varying semantic similarity thresholds to handle different types of content, as well as to reduce monetary costs and latencies of queries while providing desired responses to clients. 

GenerativeCache also includes an enhanced client for coordinating requests to multiple LLMs. The integration of our caching algorithms with the enhanced client and user interface are critically important for making the caching algorithms effective.





\bibliographystyle{ACM-Reference-Format}
\bibliography{cs_papers}

\end{document}